\documentclass[aps,prx,superscriptaddress,twocolumn,10pt]{revtex4-2}

\usepackage[utf8]{inputenc}
\usepackage[english]{babel}

\usepackage[titletoc,toc,title,page]{appendix}
\usepackage{csquotes}

\usepackage{microtype} 

\usepackage{bm} 

\usepackage{dsfont} 
\usepackage{amsmath,amssymb,amsthm,thmtools}
\usepackage{mathtools}
\usepackage{cases}
\usepackage{calc}
\usepackage{mathrsfs} 
\usepackage[normalem]{ulem} 
\newcommand\redsout{\bgroup\markoverwith{\textcolor{red}{\rule[0.5ex]{2pt}{1.pt}}}\ULon}
\usepackage{parskip}
\usepackage{nameref}
\usepackage[colorlinks=true]{hyperref}
\usepackage[nameinlink]{cleveref}
\crefname{appsec}{Appendix}{Appendices}
\crefname{box}{Box}{Box}
\hypersetup{
  colorlinks   = true, 
  urlcolor     = green!80!black, 
  linkcolor    = blue, 
  citecolor    = red!80!black 
}

\usepackage{physics} 
\usepackage{float} 
\usepackage{graphicx}
\usepackage[usenames,dvipsnames,table]{xcolor}
\usepackage{easyReview}
\usepackage{tikz}
\usetikzlibrary{calc,shapes.geometric}

\usepackage{placeins}
\usepackage{multirow,tabularx,booktabs}
\usepackage[most]{tcolorbox}
\newtcbtheorem{tbox}{Box}{enhanced, float*=t, width=\textwidth, label type=box}{box}
\usepackage[printonlyused,withpage,nohyperlinks,smaller]{acronym}
\graphicspath{{./figs/}}

\newcommand{\calH}{\mathcal{H}}

\newcommand{\bs}[1]{\boldsymbol{#1}}

\newcommand{\Strain}{\mathbf{S}^{\text{train}}}

\newcommand{\Ytrain}{\mathbf{Y}^{\text{train}}}
\newcommand{\Ytest}{\mathbf{Y}^{\text{test}}}
\newcommand{\Ypred}{\mathbf{Y}^{\text{pred}}}

\newcommand{\ytestk}{\mathbf{y}_k^{\text{test}}}
\newcommand{\ypredk}{\mathbf{y}_k^{\text{pred}}}

\renewcommand{\selectlanguage}[1]{}
\begin{document}

\title{State estimation with quantum extreme learning machines beyond the scrambling time}

\author{Marco Vetrano}
\email{marco.vetrano02@unipa.it}
\let\comma,
\affiliation{Universit\`a degli Studi di Palermo\comma{} Dipartimento di Fisica e Chimica - Emilio Segr\`e\comma{} via Archirafi 36\comma{} I-90123 Palermo\comma{} Italy}
\author{Gabriele Lo Monaco}
\let\comma,
\affiliation{Universit\`a degli Studi di Palermo\comma{} Dipartimento di Fisica e Chimica - Emilio Segr\`e\comma{} via Archirafi 36\comma{} I-90123 Palermo\comma{} Italy}
\author{Luca Innocenti}
\let\comma,
\affiliation{Universit\`a degli Studi di Palermo\comma{} Dipartimento di Fisica e Chimica - Emilio Segr\`e\comma{} via Archirafi 36\comma{} I-90123 Palermo\comma{} Italy}
\author{Salvatore Lorenzo}
\affiliation{Universit\`a degli Studi di Palermo\comma{} Dipartimento di Fisica e Chimica - Emilio Segr\`e\comma{} via Archirafi 36\comma{} I-90123 Palermo\comma{} Italy}
\author{G. Massimo Palma}
\affiliation{Universit\`a degli Studi di Palermo\comma{} Dipartimento di Fisica e Chimica - Emilio Segr\`e\comma{} via Archirafi 36\comma{} I-90123 Palermo\comma{} Italy}

\begin{abstract}
    Quantum extreme learning machines (QELMs) leverage untrained quantum dynamics to efficiently process information encoded in input quantum states, avoiding the high computational cost of training more complicated nonlinear models.
    On the other hand, quantum information scrambling (QIS) quantifies how the spread of quantum information into correlations makes it irretrievable from local measurements.
    Here, we explore the tight relation between QIS and the predictive power of QELMs.
    In particular, we show efficient state estimation is possible even beyond the scrambling time, for many different types of dynamics --- in fact, we show that in all the cases we studied, the reconstruction efficiency at long interaction times matches the optimal one offered by random global unitary dynamics.
    These results offer promising venues for robust experimental QELM-based state estimation protocols, as well as providing novel insights into the nature of QIS from a state estimation perspective.
\end{abstract}

\maketitle
\section{Introduction}
\label{Intro}

The capability of complex systems to store, propagate, and process information is at the core of reservoir computing (RC)~\cite{lukosevicius_reservoir_2009,lukosevicius_practical_2012,angelatos_reservoir_2021,jaeger_harnessing_2004} and extreme learning machines (ELM)~\cite{konkoli_reservoir_2017, huang_extreme_2004,huang_extreme_2011, wang_review_2022, markowska-kaczmar_extreme_2021}.
RCs and ELMs are supervised machine learning techniques that leverage a complex unknown fixed dynamic, referred to as ``reservoir'' in this context, to quickly process data in order to make target features easier to retrieve via a simple linear regression, with RCs also capable of processing temporal data thanks to their use of reservoirs with memory.
The simplicity of the training and the heterogeneity of systems that can be used as reservoirs~\cite{soriano_delay-based_2015, bhovad_physical_2021, nakajima_information_2015, coulombe_computing_2017, goudarzi_dna_2013, tanaka_recent_2019, nokkala_high-performance_2022}, are at the core of the success of ELMs.
Classical ELMs (RCs) are quantized replacing classical functions with physical quantum dynamics, giving rise to so-called \textit{quantum} extreme learning machines (QELMs) and \textit{quantum} reservoir computing (QRCs)~\cite{mujal_opportunities_2021, govia_quantum_2021, martinez-pena_quantum_2023, ghosh_quantum_2019,ghosh_quantum_2019-1, martinez-pena_dynamical_2021, ghosh_realising_2021, s_ghosh_reconstructing_2021, mujal_time-series_2023, krisnanda_creating_2021,garcia-beni_scalable_2023,martinez-pena_information_2020, xiong2023fundamental, vcindrak2024enhancing, vcindrak2024krylov}. QELMs, among other things, have shown significant promise for experimental state estimation tasks~\cite{innocenti_potential_2023,suprano_experimental_2023,s_ghosh_reconstructing_2021}.
In this context, the fact that information is most often collected from \textit{local} measurements on the output states, naturally raises questions about the relations between QELM-based state estimation protocols and the spreading of information throughout its reservoir. 

On the other hand, quantum information scrambling (QIS)~\cite{hayden_black_2007, zhuang_scrambling_2019, shenker_black_2014,landsman_verified_2019, lashkari_towards_2013, roberts2015localized, sekino_fast_2008, hosur_chaos_2016, lo_monaco_quantum_2023, yan_information_2020, zanardi2021information,Campisi_2017,swingle_measuring_2016,xu_scrambling_2024,swingle2018unscrambling, gherardini2024quasiprobabilities} is a framework to investigate the retrievability of information from local measurements on output states. 
Dynamics are said to be \textit{scrambling} when they hide information in the internal correlations of the system, thus rendering it irretrievable through local measurements~\cite{touil2020quantum, touil2024information}.
Several quantifiers have been proposed to study this phenomenon, including out-of-time-ordered-correlators (OTOCs)~\cite{swingle_measuring_2016, hosur_chaos_2016, zanardi2021information, hashimoto_out--time-order_2017}, tripartite information~\cite{hosur_chaos_2016,lo_monaco_quantum_2023,lomonaco2023operational,schnaack_tripartite_2019}, and generalized channel capacities~\cite{touil2020quantum,yan_information_2020,yuan_quantum_2022,zhuang2022phase,lomonaco2023operational}.

{
Since QELM-based architectures commonly rely on the information retrieved by measuring a set of local observables on the reservoir, understanding how QIS affects information retrievable from such measurements provides valuable insights, which can guide the search for dynamics suitable for experimental implementations.
Furthermore, given the nature of QIS, one might expect that because of scrambling, performing task via local measurements beyond the \textit{scrambling time} --- the time after which OTOCs saturate to their asymptotic value~\cite{swingle_measuring_2016, shenker_black_2014, maldacena2016bound, hosur_chaos_2016} --- would be highly inefficient due to the loss of information in correlations.
On the other hand, previous studies~\cite{martinez-pena_dynamical_2021,xia2022reservoir} have pointed out that systems exhibiting ergodic to non-ergodic phase transitions influence both the information processing capabilities and short-term memory of reservoirs in the context of QRC. These studies also show a connection to the scrambling capabilities of the system. Deepening our understanding of QIS in local-measurements-based QELM can thus be a resource also in the context of QRC, helping to search for reservoir topologies that maximize the information retrieved with a weak measurement approach~\cite{mujal_time-series_2023}.

In this work, we investigate the interplay between QELM-based state estimation and QIS. We find, in particular, that information remains consistently retrievable from local measurements even far beyond the scrambling time.
This observation highlights that the crucial property for state reconstruction is information spreading, rather than information scrambling \textit{per se}.
}
These results run counter to the standard notion that beyond the scrambling time, all information about input states is lost into non-local correlations.
Instead, we observe that even when OTOCs would indicate the presence of scrambling, the amount of residual local information is sufficient to reconstruct input states efficiently, and remains so even at longer timescales.
These results have two significant implications.
First, they reveal that analyzing scrambling systems through the lens of QELM-based state estimation offers new insights into the nature of QIS and its quantifiers.
Second, they suggest that even scrambling systems can support efficient and robust input state reconstruction, providing experimentally viable platforms for quantum state estimation.

Furthermore, we observe two different regimes characterized by different behaviours of the estimation accuracy as a function of time.
In a first transient regime, characterized by the OTOC linearly increasing with time --- which signals that information is still spreading through the system --- the estimation accuracy depends on the reservoir structure, and is reflected by the behaviour of the Holevo information.
On the other hand, beyond the scrambling time --- defined by the saturation of the OTOC --- these differences disappear, and the estimation accuracy stabilizes to a constant value.
We thus find while OTOCs correspond to the scrambling time of the system, entropic quantifiers provide a more fine-grained insight into the estimation accuracy of QELMs. 

{The article is organized as follows: in  \cref{sec:ELM} and~\cref{sec:scrambling} we briefly review ELMs, QELMs, and QIS.
In~\cref{sub:reservoir_dimension} and~\cref{sub_:estimation} we present our main results, while we draw our conclusions in~\cref{sec:conclusions}. 
Finally, in~\cref{sec:task} we include more details about the methods used throughout the paper.}



\section{Results}
\label{sec:results}

\subsection{Review of ELMs and QELMs}
\label{sec:ELM}

From a broad mathematical perspective, given a parametrized family of functions $f_{\bs\theta}$, the goal of a machine learning algorithm is to ``learn'' the values of the parameters $\bs\theta$ that implement a target relation between input and output data. The function $f_{\bs\theta}$ is referred to as the ``model'' in this context, and $\bs\theta$ are the parameters to train.
In particular, for \textit{supervised} machine learning models, a \textit{training dataset} $(\mathbf S,\mathbf Y)\equiv \{(\mathbf s_k,\mathbf y_k)\}_{k=1}^{N_{\rm train} }$, with $\mathbf s_k\in\mathbb{R}^{N_{\rm in}}$ and $\mathbf y_k\in\mathbb{R}^{N_{\rm feat}}$, is used to find $\bs\theta$ corresponding to which $f_{\bs\theta}(\mathbf s_k)\simeq \mathbf y_k$ for all $k$.
Here $N_{\rm train},N_{\rm in}$, and $N_{\rm feat}$ are the number of training vectors, input vectors, and features, respectively.
To quantify how well a given model is performing, we define a loss function $L(f(\mathbf s_k),\mathbf y_k)$, and seek to minimize its expectation value over the training dataset, and then test its performance on a testing dataset of previously unseen data.
A common choice of such loss function is the standard Euclidean distance, whose expectation value is then just the mean-squared error (MSE)~\cite{goodfellow_deep_2016}.
The optimization of the model is then typically, but not exclusively, performed using stochastic-gradient-descent-based methods, which iteratively tune $\bs\theta$ to minimize the loss~\cite{ruder_overview_2017}.

\subsubsection{ELMs}

An ELM is a particularly simple type of machine learning model which involves a complex but untrained function $R:\mathbb{R}^{N_{\rm in}}\to\mathbb{R}^{N_{\rm out}}$, referred to as the \textit{reservoir function}, followed by a trained linear layer $\mathbf W$, so that the overall model can be formally written as
    $f^{\text{ELM}}_{\bf W}=\mathbf W\circ R$.
The reservoir function is often implemented as a recurrent neural network with randomly initialized weights~\cite{lukosevicius_reservoir_2009}.
The advantage of this model is to reap the generalization benefits of highly nonlinear functions, while at the same time avoiding the high computational cost of training them.

Training an ELM thus amounts to solving a linear regression problem, which is generally much simpler than training a traditional deep neural network~\cite{lukosevicius_reservoir_2009}.
More precisely, given a training dataset $(\Strain,\Ytrain)$, we want to find an exact or approximate $\mathbf W$ such that
\begin{equation}\label{eq:linear_problem_training}
    \mathbf W R(\Strain) =\mathbf Y^{\rm train},
\end{equation}
where $R(\Strain)$ is the matrix whose $k$-th column is $R(\mathbf s_k^{\rm train})$.
Here, $R(\Strain)$ is an $N_{\rm out}\times N_{\rm train}$ matrix, while $\mathbf Y^{\rm train}$ is $N_{\rm feat}\times N_{\rm train}$.
The least squares solution to~\cref{eq:linear_problem_training} is
\begin{equation}\label{eq:classical_training}
\mathbf{W}=\Ytrain R(\Strain)^+,
\end{equation}
where $R(\Strain)^+$ denotes the Moore-Penrose pseudo-inverse, which can be computed via the singular value decomposition (SVD) as $R(\Strain)^+ = U\Sigma^+ V^\dagger$, if $R(\Strain)=U\Sigma V^\dagger$ is the SVD of the original matrix, with $U,V$ isometries and $\Sigma>0$ positive diagonal squared matrix with the singular values as diagonal elements~\cite{serre2010matrices}.

Although linear systems are generally relatively easy to solve, there are many situations where the solution can be numerically unstable.
A common way to quantify the stability of the solution for $\mathbf A$ of a linear system $\mathbf{Y} = \mathbf{AX}$, with $\mathbf X,\mathbf Y$ matrices, is the condition number $\kappa(\mathbf{X})$ defined as~\cite{belsley2005regression}
\begin{equation}
    \kappa(\mathbf{X}) = \frac{\abs{\lambda_{\rm max}(\mathbf{X})}}{\abs{\lambda_{\rm min}(\mathbf{X})}}
    \label{eq: condition_number}
\end{equation}
where $\lambda_{\rm max}$ and $\lambda_{\rm min}$ are the maximal and minimal singular values of $\mathbf{X}$.
The condition number quantifies how perturbations in $\mathbf X$ result in errors in the estimated $\mathbf A$. High condition numbers, typically corresponding to a near-singular $\mathbf X$, indicate that small errors in $\mathbf X$ could result in large errors in $\mathbf A$, and thus that the linear system is ill-conditioned.

\subsubsection{QELMs}
\label{sec:QELMs}

In QELMs, the vectors $\mathbf s_k$ become input quantum states $\rho_k$, and the reservoir function $R$ is replaced by the combination of a quantum dynamic $\Phi$ and a measurement~\cite{mujal_opportunities_2021,innocenti_potential_2023}.
The measurement can be modelled as either some POVM or, as we will do here, some set of observables $\{\mathbf O_j\}_{j=1}^{N_{\rm out}}$.
The classical matrix $R(\Strain)$ becomes for QELMs the matrix of probabilities $\mathbf P^{\rm train}$ defined as
\begin{equation}
    [\mathbf P^{\rm train}]_{jk} \equiv \Tr[ \mathbf O_j \Phi(\rho_k^{\rm train})],
    \label{eq: qelm}
\end{equation}
with $\{\rho_k^{\rm train}\}_{k=1}^{N_{\rm train}}$ the states in the training dataset. The matrix $\mathbf P^{\rm test}$ is defined analogously from the testing states $\{\rho_k^{\rm test}\}_{k=1}^{N_{\rm test}}$.
Note that $\mathbf P^{\rm train}$ has dimensions $N_{\rm out}\times N_{\rm train}$ with $N_{\rm out}$ the number of measurement outcomes and $N_{\rm train}$ the number of training states.
Training the QELM involves again solving the linear system~\eqref{eq:classical_training}, upon replacing
$R(\Strain) \to \mathbf P^{\rm train}$.
The matrix $\Ytrain$ in the quantum case contains the set of $N_{\rm feat}$ features associated to each of the $N_{\rm train}$ training input states.
We will focus on the case where the target features are expectation values of target observables, meaning that $[\mathbf Y^{\rm train}]_{ij} = \Tr[\mathbf O_i \Phi(\rho_j^{\rm train})]$ with $\mathbf O_i$ the $i=1,..., N_{\rm feat}$ labelling the target observables, and similarly for $\Ytest$.

The training process produces a linear map $\mathbf{W}$ which applied to measurement probabilities recovers the target observables.
To evaluate the performance of the trained reservoir we use $\mathbf P^{\rm test}$ to compute the predicted features
$\Ypred = \mathbf W \mathbf P^{\rm test}$, and compute the MSE as
\begin{equation}
     \text{MSE}(\Ytest,\Ypred)=\frac{1}{N}\sum_{k=1}^N \vert \ytestk-\ypredk|^2.
     \label{eq: quant_MSE}
\end{equation}

The output expectation values can always be written as $\Ytrain_{ij} = \Tr[\Phi^\dagger(\mathbf O_i)\rho_j^{\rm train}]$ with $\Phi^\dagger$ the dual of $\Phi$. This amounts to describing measurements in the Heisenberg pictures, and allows to concisely describe QELMs as involving a direct measurement on input state. It follows that a target observable $\mathbf O$ can be accurately estimated iff it can be written as a linear combination of the operators $\Phi^\dagger(\mathbf O_j)$~\cite{innocenti_potential_2023}.

\subsection{Review of QIS}
\label{sec:scrambling}

Quantum information scrambling (QIS) studies the spread of local information across a many-body system, and its retrievability from local measurements~\cite{hayden_black_2007, zhuang_scrambling_2019, shenker_black_2014,landsman_verified_2019, lashkari_towards_2013, roberts2015localized, sekino_fast_2008, hosur_chaos_2016, lo_monaco_quantum_2023, yan_information_2020, zanardi2021information,Campisi_2017,swingle_measuring_2016,xu_scrambling_2024, gherardini2024quasiprobabilities}.
Two main approaches to quantify QIS are out-of-time-ordered-correlators (OTOCs) \cite{hashimoto_out--time-order_2017, swingle2018unscrambling, swingle_measuring_2016, xu_scrambling_2024, hosur_chaos_2016, zanardi2021information, gherardini2024quasiprobabilities}, and entropic quantifiers such as Holevo information~\cite{yuan_quantum_2022, zhuang2022phase} and tripartite information~\cite{hosur_chaos_2016,lomonaco2023operational,lo_monaco_quantum_2023}.
We focus on the first two, due to them being easier to compute and relate to the estimation accuracy of QELMs.
In this section, we will briefly review these two methods, in order to set the stage for the next sections.

\subsubsection{OTOCs}
\label{section:OTOC}

The idea behind OTOCs~\cite{hashimoto_out--time-order_2017, xu_scrambling_2024, swingle_measuring_2016, swingle2018unscrambling, hosur_chaos_2016, zanardi2021information, gherardini2024quasiprobabilities} is to quantify QIS via the non-commutativity of observables at different times.
OTOCs have been measured experimentally on digital quantum computers based on trapped ions and on nuclear magnetic resonance quantum simulators~\cite{green_experimental_2022, li_measuring_2017, nie2020experimental,halpern2018quasiprobability}.

Consider two disjoint subsystems $\mathcal H_A$ and $\mathcal H_B$ of a larger Hilbert space $\mathcal H$ and suppose our goal is to encode information in $\calH_B$ and retrieve it from $\calH_A$ after a time $t$, while the overall system undergoes a unitary evolution $U$.
OTOCs quantify the viability of this process through the correlator between $\mathbf O_B$ and the Heisenberg-evolved $\mathbf O_A(t) \equiv \Phi_t^\dagger(\mathbf O_A)$, for different pairs of local operators $\mathbf O_A$ and $\mathbf O_B$, with $\Phi_t$ the dynamical map describing the evolution, and $\Phi_t^\dagger$ its adjoint.
In the case of unitary evolution, this simplifies to $\Phi_t(\rho)= U_t \rho U_t^\dagger$ with $U_t=e^{-iHt}$, and thus $\Phi_t^\dagger(\mathbf O_A)=U_t^\dagger \mathbf O_A U_t$.
The non-commutativity is quantified as~\cite{zanardi2021information}
\begin{equation}
     C(t) = \frac{1}{2d}\Tr{[\mathbf{O}_A(t), \mathbf{O}_B]^\dagger[\mathbf{O}_A(t), \mathbf{O}_B]}.
     \label{eq: C}
\end{equation} 
When $\mathbf{O}_A$ and $\mathbf{O}_B$ are both Hermitian and unitary,~\cref{eq: C} simplifies to~\cite{jalabert_semiclassical_2018, zanardi2021information}
\begin{equation}
    C(t) = 1 - \frac{1}{d}\Tr{\mathbf{O}_A(t)\mathbf{O}_B\mathbf{O}_A(t)\mathbf{O}_B}.
    \label{eq: C otoc}
\end{equation}
To relate OTOCs to QELMs, we will take $\mathbf O_A,\mathbf O_B$ to be Pauli operators.

\subsubsection{Holevo information}
\label{sec:holevo}

The Holevo information between input and individual output states has been shown to be a viable alternative quantifier for QIS~\cite{holevo1973bounds,yuan_quantum_2022,zhuang2022phase}.

In general, the Holevo information provides an upper bound to the accessible correlations between a sender and a receiver, in situations where quantum states are used as a medium for classical information transmission.
More specifically, suppose Alice sends a classical message to Bob, encoding it into the choice of a state taken from an ensemble $\eta\equiv\{(p_i,\rho^i)\}_{i=1}^M$ for some integer $M$. Bob wants to recover the message from the measurement results on the state he received. Then, the appropriate quantifier for the correlations between Alice and Bob is the \textit{accessible} mutual information $I_{\rm acc}(A:B)$~\cite{holevo1973bounds,qi2022holevo,nielsen_quantum_2010}.
This is defined as the classical mutual information of the joint probability distribution
\begin{equation}
    p_{ik}(\bs\Pi) \equiv p_i \Tr(\Pi_k \rho^i),
\end{equation}
maximized over all POVMs $\bs\Pi\equiv\{\Pi_k\}_k$ that Bob can perform on the state he receives.
Further maximizing $I_{\rm acc}(A:B)$ over all possible ensembles $\eta$, gives the classical capacity of the channel representing the dynamics~\cite{watrous2018theory}.
However, the accessible information is often hard to compute, due to the optimization involved in its definition.
Nonetheless, Holevo's theorem upper bounds $I_{\rm acc}(A:B)$ in terms of the \textit{Holevo information} $\chi(\eta)$, whose computation notably does not require to perform an optimization~\cite{nielsen_quantum_2010,watrous2018theory}:
\begin{equation}
\begin{gathered}
    I_{\rm acc}(A:B)\le \chi(\eta), \\
    \chi(\eta) = S\left( \sum_i p_i \rho^i \right) -
    \sum_i p_i S(\rho^i),
\end{gathered}
\end{equation}
where $S(\rho)$ denotes the von Neuamm entropy of $\rho$.

To relate QIS to the estimation accuracy of QELMs we want to study how the information accessible \textit{locally} from the output qubits evolves over time.
Let us then consider the dynamical maps $\Phi_t^{(j)}$ which describe the evolution up to time $t$, followed by a partial trace over all but the $j$-th output qubit.
Denoting with
\begin{equation}
    \Phi_t^{(j)}(\eta) \equiv
    \{(p_i, \Phi_t^{(j)}(\rho^i))\}_{i=1}^M
\end{equation}
the ensemble obtained applying $\Phi_t^{(j)}$ to each state in $\eta$, the \textit{local Holevo informations} $\chi(\Phi_t^{(j)}(\eta))$ then quantify the information recoverable from the $j$-th output qubit, at time $t$, for each reservoir qubits $j=1,..., N$. 
{The channel $\Phi_t^{(j)}$ thus captures how input states evolve into the $j$-th marginal reservoir qubit over time, with the Holevo information $\chi(\Phi^{(j)}_t(\eta))$ providing an upper bound on the information about the input states that can be recovered by measuring that reservoir qubit.
The connections between mutual information, Holevo $\chi$, and QIS, have been further explored in~\cite{lo_monaco_quantum_2023,lomonaco2023operational,yuan_quantum_2022}.
In particular, the extreme values of the Holevo $\chi$ reflect into deeply different behaviours of the channel: namely, when the Holevo $\chi$ vanishes, input states become indistinguishable at the output, so that no useful information can be extracted.
By contrast, for larger values of $\chi$, there are measurement choices that allow to distinguish different input states with minimum error rate.
}

Following~\cite{yuan_quantum_2022}, we consider the Holevo information with unbalanced prior probabilities $p_i=1/M$.
Using unbiased priors more closely relates to the task of parameter estimation in QELMs, where the estimation is performed without assuming any prior knowledge of the input states.

\begin{figure}
    \centering
    \includegraphics[width=\linewidth]{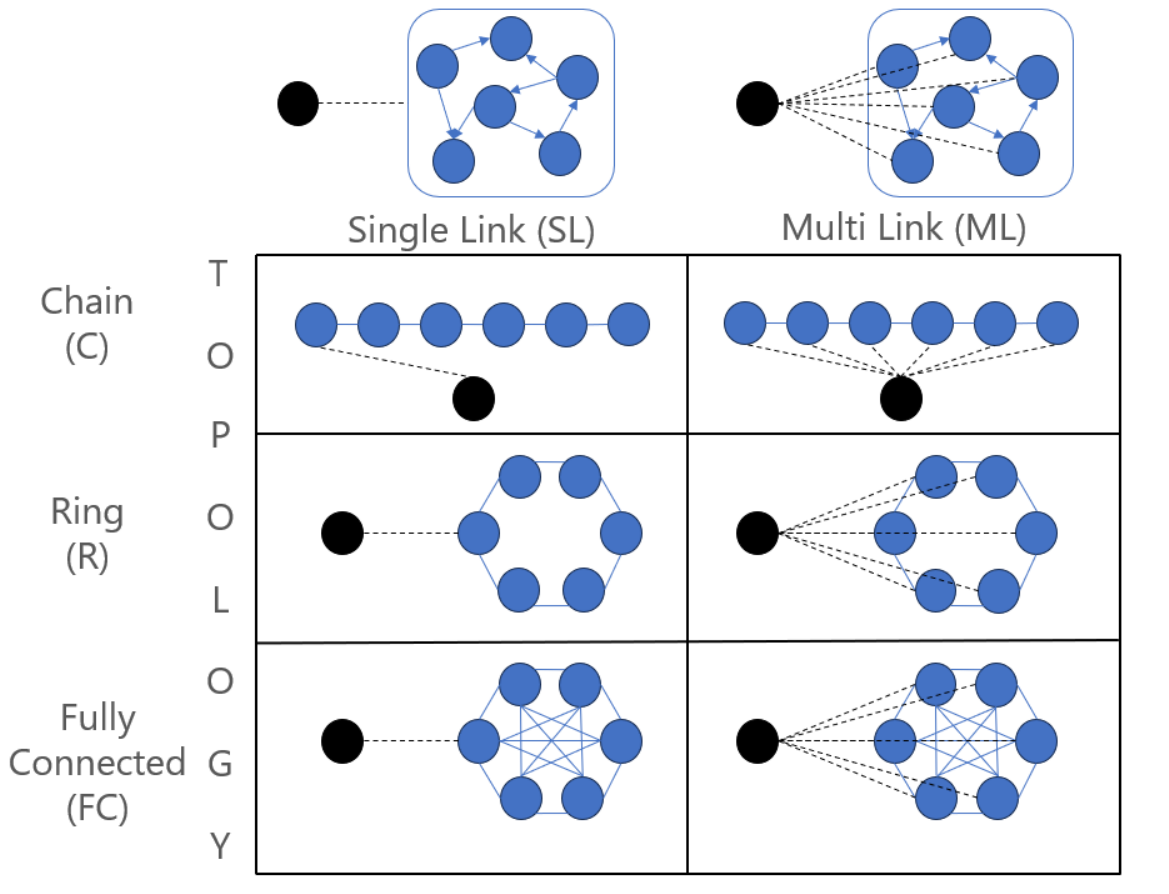}
    \caption{
    Summary of interaction topologies and input couplings used throughout the paper.
    The different interaction topologies, \textit{chain} (C), \textit{ring} (R), and \textit{fully connected} (FC) correspond to different Hamiltonian terms $H_{\rm res}$, while the two input couplings, \textit{single link} (SL) and \textit{multi-link} (ML) to different Hamiltonian terms $H_{\rm inj}$.
    The dashed black lines in the figures represent interaction terms between input and reservoir qubits, while the solid blue lines represent interaction terms between reservoir qubits.
    }
    \label{fig:configurazioni}
\end{figure}

\begin{figure}[t]
    \centering
        \includegraphics[width = 0.9\linewidth]{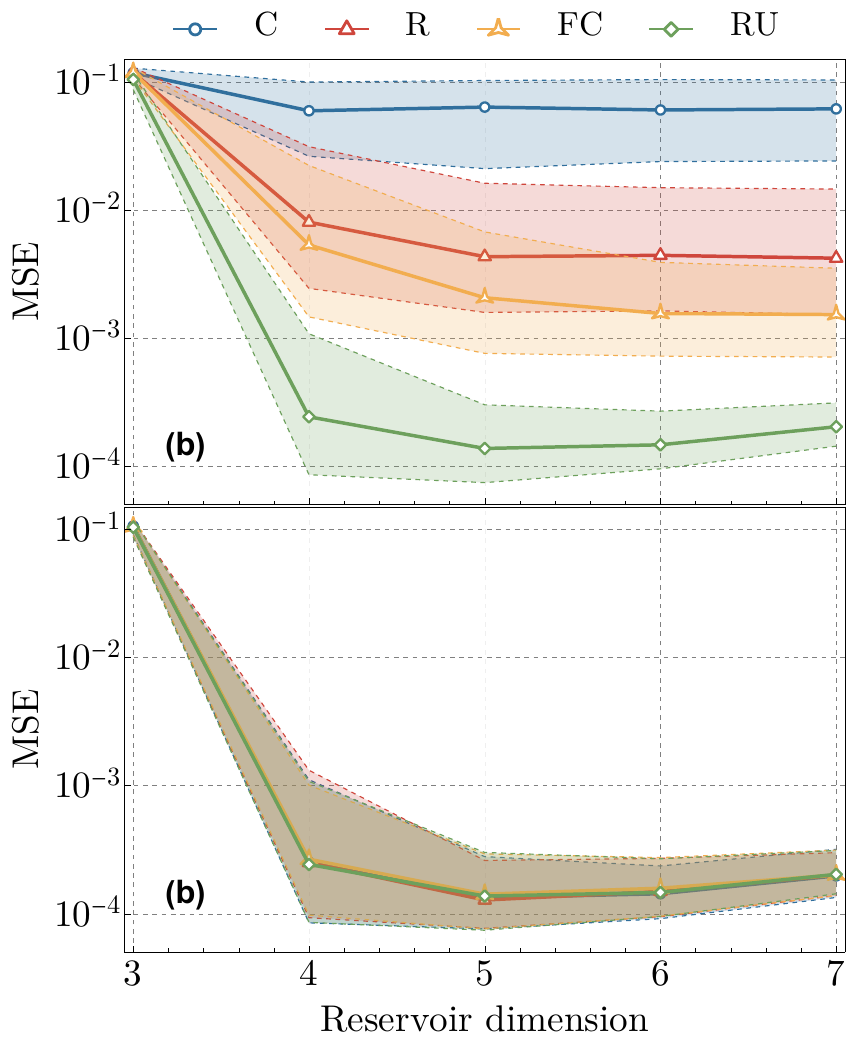}
    \caption{Reconstruction MSE vs number of qubits in the reservoir, with SL input coupling and different reservoir topologies. The evolution time is {\bf(a)} $t=0.25$ and {\bf(b)} $t=5$.
    Each point represents the median of the MSEs computed on an ensemble of $500$ random Hamiltonians, with sampling statistics of $10^6$ for both training and testing in each case. Training and testing sets were each comprised of $50$ random states.
    Errors bars represent first and third quartiles of the data.
    In addition to the standard topologies C, R, FC defined in~\cref{fig:configurazioni}, we also report the MSE obtained letting the whole input and reservoir system evolve unitarily with a Haar random unitary (RU).
    }
    \label{fig:var_res_dim}
\end{figure}

It is to be expected that the local Holevo information is closely related to the reconstruction performances of QELMs, as it quantifies the maximum achievable asymptotic transmission rate when maximized over input ensembles~\cite{watrous2018theory}.
While we do not perform this optimization explicitly, there is a clear link between achievable transmission rates and state reconstruction efficiency.
Consequently, the Holevo information serves as an effective and easily computable measure of correlations for our purposes.
{
Namely, the Holevo information provides a straightforward quantifier of the correlations between input states and local output states, serving as a natural indicator of the information about the input state recoverable from the outputs.
}

\subsection{Reservoir dimension}
\label{sub:reservoir_dimension}
The estimation accuracy of QELMs was previously shown to depend on the reservoir dynamics~\cite{innocenti_potential_2023,suprano_experimental_2023, ghosh_realising_2021}, and in particular, for time-independent Hamiltonian dynamics, on the topology of interactions~\cite{innocenti_potential_2023}.
However, as we will show, if systems are left to evolve beyond the scrambling time most of these differences disappear.
We find that in all the cases we studied, reconstruction remains possible beyond the scrambling time with a good level of accuracy without the need to fine-tune the interaction time.
Furthermore, we find that the Holevo information correlates with the reconstruction MSE even where the OTOC does not, highlighting a difference in the QIS features they measure.

\begin{figure}[tbh]
\centering
\hspace*{-1cm}
    \includegraphics[width=1.12\linewidth]{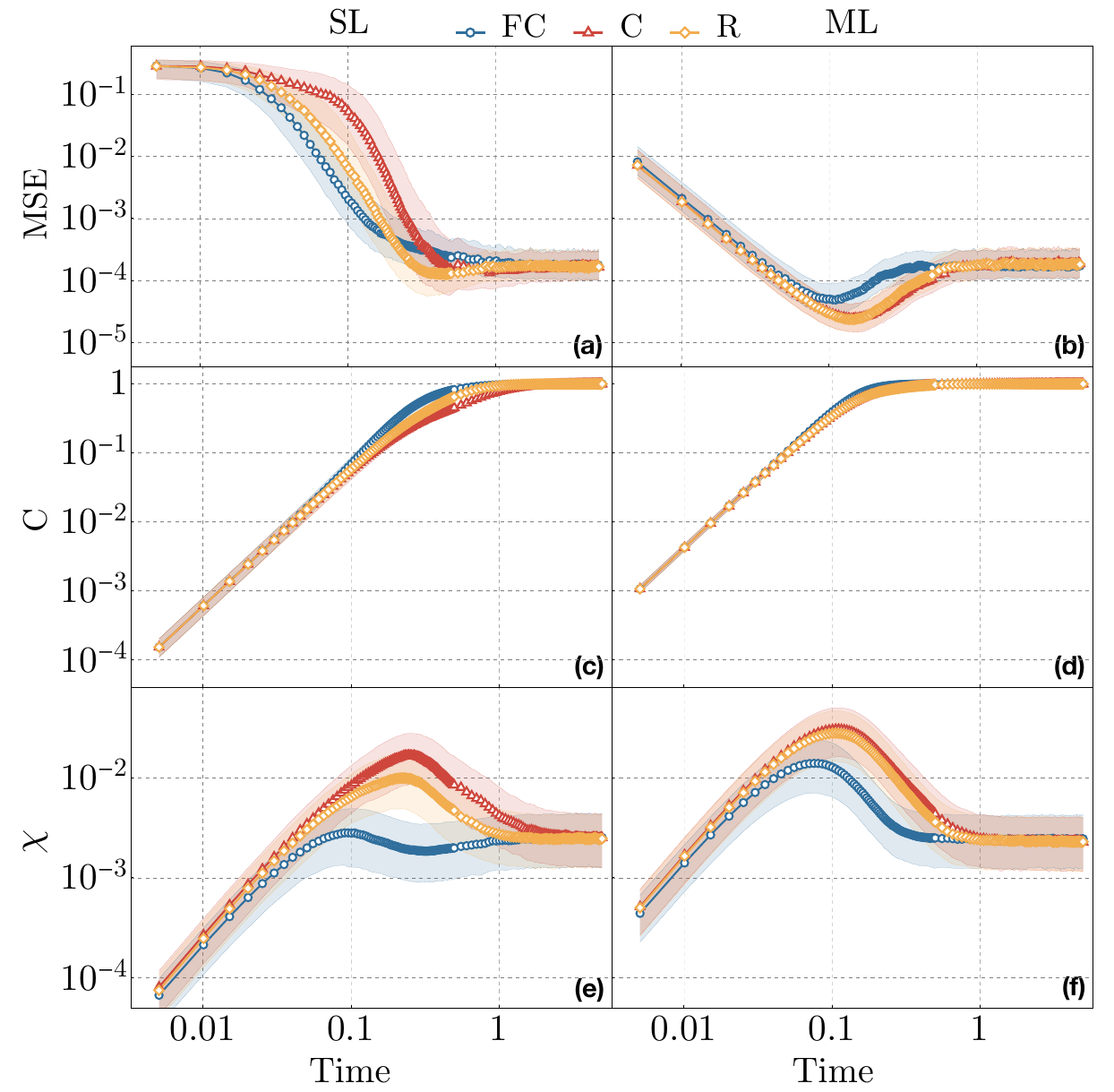}
    \caption{Reconstruction MSE {\bf (a)}-{\bf(b)}, two-point correlation function $C$ {\bf (c)}-{\bf(d)}, and local Holevo information $\chi$ {\bf (e)}-{\bf(f)}, as a function of time, for different interaction topologies, as a function of time in the interval $t\in[0,5]$. In each plot we present the data corresponding to the three interaction topologies, FC (blue circles), C (red triangles), R (orange diamonds), outlined in~\cref{fig:configurazioni}. The left realizations {\bf (a)}-{\bf (c)}-{\bf (e)} refer to the SL scheme, while the right realizations {\bf (b)}-{\bf (d)}-{\bf (f)} the ML scheme. 
    In each case, we present the median results over 500 realizations of random Hamiltonians with the corresponding topology, with a reservoir of $N=7$ qubits plus a single input qubit. The error bars show the first and third quartiles around the median.
    For the MSE we used sampling statistics of $10^6$ in both training and test, both of which were performed with training and testing dataset comprised of $50$ random states each.
    }
    \label{fig:General Results}
\end{figure}

\begin{figure}[tbh]
    \centering
    \includegraphics[width=0.85\linewidth]{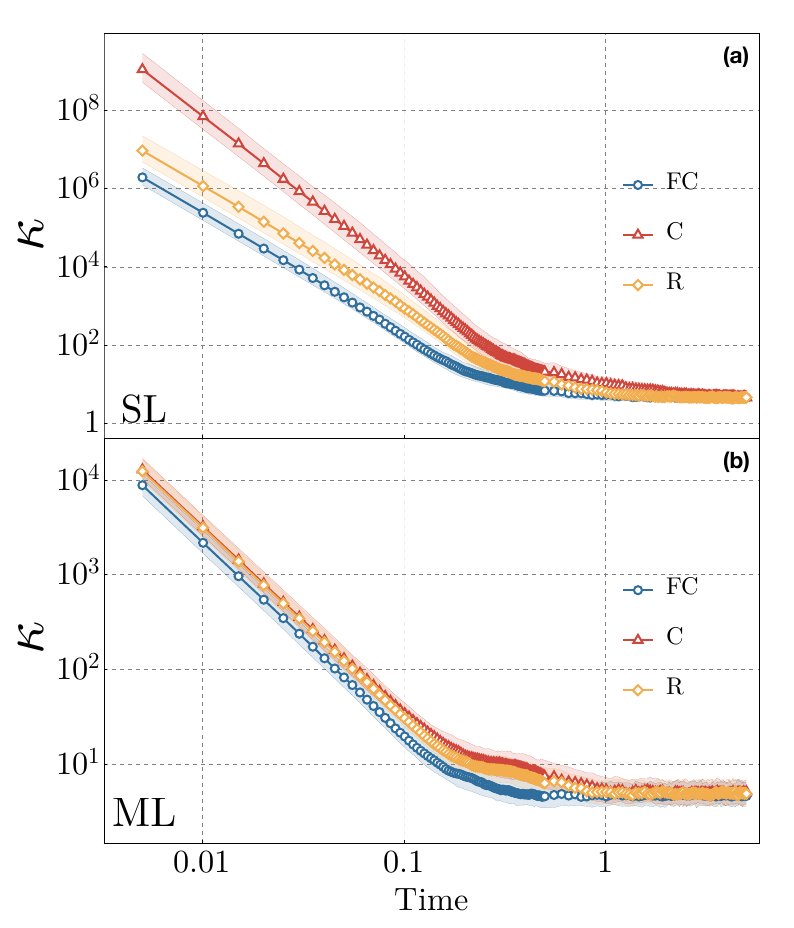}
  \caption{Condition number of $\mathbf P^{\rm train}$ for the topologies of \cref{fig:configurazioni} with SL ($\bf{a}$) and ML ($\bf{b}$) coupling schemes, for different evolution times $t\in[0,5]$.
  In each topology and coupling scheme, each point gives to the condition number averaged over 500 random Hamiltonians, a reservoir with $N=7$ qubits, and sampling statistics of $10^6$ for both training and testing.
  The error bars represent the first and third quartiles. Training and testing were conducted with 50 states each.
  The long-time condition numbers are $\kappa = 4.6\pm1.1$ and $\kappa = 5.1\pm1.2$ for SL and ML schemes, respectively.
  }
  \label{fig:CN}
\end{figure}

\begin{figure*}
  \centering
  \includegraphics[width=\linewidth]{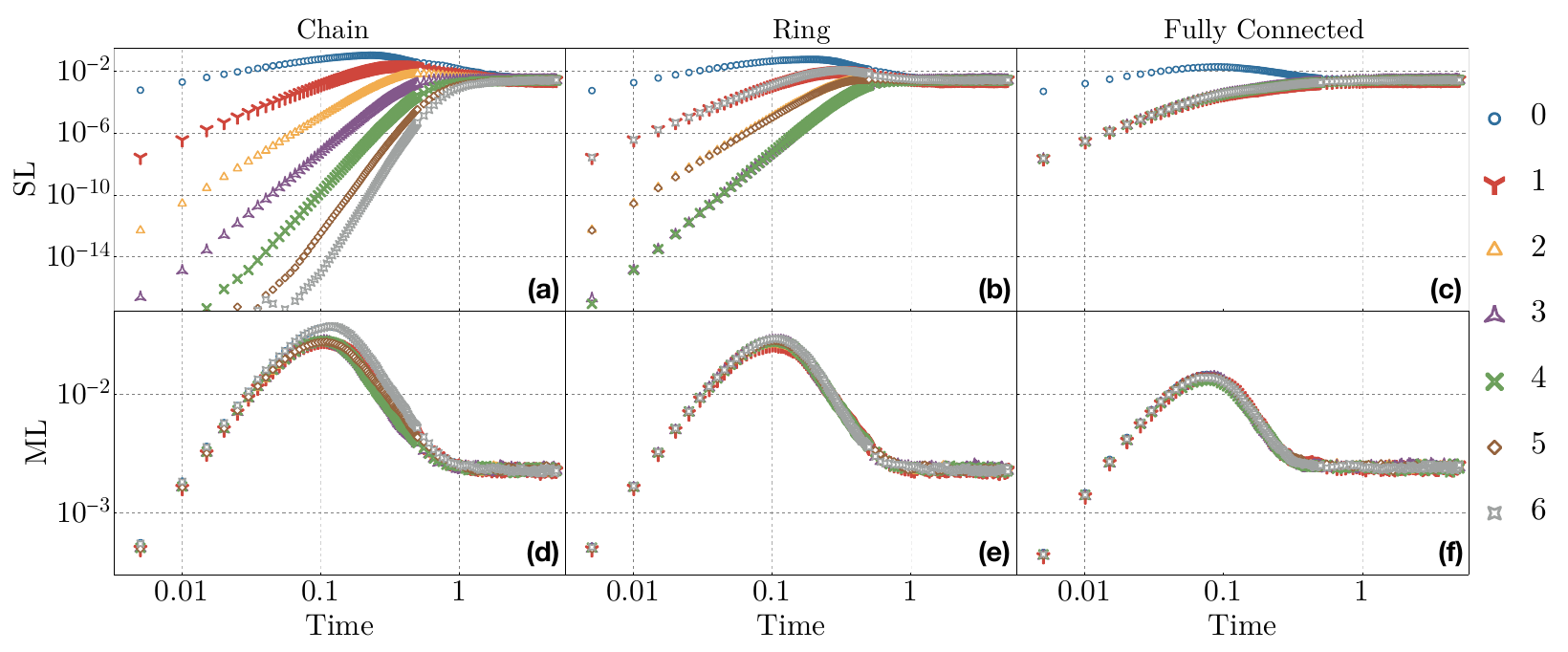}
  \caption{Distribution of the Holevo information among the nodes of the reservoir for SL (\textbf{a,b,c}) and ML (\textbf{d,e,f}) input coupling, and for chain (\textbf{a,d}), ring (\textbf{b,e}), and fully connected (\textbf{c,f}) interaction topologies.
  Each figure reports the Holevo $\chi$ corresponding to each of the $N=7$ reservoir qubits, here labelled from $0$ to $6$, for different evolution times $t\in[0,5]$.
  Each data point is the median of the dataset obtained over an ensemble of $500$ random Hamiltonians.}
  \label{fig:holevo single nodes}
\end{figure*}

In~\cref{fig:var_res_dim} we report the reconstruction MSE as a function of the reservoir dimension --- from $2$ to $7$ qubits --- for the different interaction topologies, with the SL input coupling, for short ($t=0.25$) and longer ($t=5$) interaction times.
This choice of injection scheme ensures a constant number of injection links while increasing the dimension of the reservoir.
Note that when the reservoir contains less than $4$ qubits there is not enough information to retrieve the input state, which explains the higher MSEs for those points.
The data shows that although at short times we reproduce the distinction between the different interaction topologies reported in~\cite{innocenti_potential_2023}, for longer times these differences disappear.
This suggests that the propagation of information through the reservoir relatively quickly compensates and cancels out possible initial differences in the structure of interactions.
These findings are further corroborated by studying the reconstruction MSE corresponding to instead letting the whole input+reservoir system evolve unitarily. As shown in the figure, if instead of random Hamiltonian evolution we consider Haar random unitary evolutions, which generally result in highly correlated systems, we get the same results given by different Hamiltonians at long evolution times.
This suggests that the long-time behaviour of generic types of Hamiltonians approaches the behaviour given by Haar random unitaries, which are known to be optimal to estimate arbitrary observables of input states~\cite{innocenti_potential_2023,innocenti2023shadow,huang2020predicting}
\subsection{Estimation accuracy and QIS}
\label{sub_:estimation}

We then study in~\cref{fig:General Results} the time-dependence of the MSE, and relate it with OTOC and local Holevo information.
Doing so reveals that the different topologies converge to the same behaviour for evolution times longer than the scrambling time, as defined by the saturating OTOC.
This data also shows that simply studying the OTOC is not sufficient to predict the reconstruction MSE, as the OTOC always monotonically increases until saturation.
On the other hand, the Holevo information more closely follows the MSE, as expected from it quantifying correlations between inputs and local output qubits.
In all topologies, the MSE eventually saturates to an asymptotic value of $(1.8\pm0.5)\cross10^{-4}$, the averaged $\chi$ to $(2.5\pm1.2)\times 10^{-3}$, and the OTOC to $C = 0.997\pm0.003$.
We also note how having more interaction terms in the network shortens the scrambling time, and thus how long it takes the system to reach the saturation regime. This can be traced back to the additional interaction terms allowing information to spread faster throughout the reservoir.
In~\cref{fig:CN} we report the time-dependence of the condition number $\kappa$ for all different topologies.
The condition number quantifies the degree to which relative stochastic errors are amplified in the testing stage, and is another easily computable quantity that gives insight into the numerical stability of the linear problem associated with a QELM, and therefore its associated estimation performances.

A phenomenon characteristic of ML-coupled systems seen in~\cref{fig:General Results} is a transient regime where reconstruction accuracy is better than its asymptotic value, which is also reflected in higher values of the average $\chi$ corresponding to the same evolution times.
This is due to the direct couplings in ML systems allowing information to initially spread and become locally retrievable from all reservoir qubits, to then become partially lost to the correlation at longer times, as can be seen from~\cref{fig:holevo single nodes}~\textbf{(d)-(e)-(f)}.
Furthermore, for both injection schemes, right before the saturation, C and R topologies counter-intuitively achieve somewhat better accuracies than the FC one --- this can be traced back to more information being accessible locally for these schemes.

A feature worth pointing out in the results of~\cref{fig:General Results} is how in some cases the average $\chi$ seems to contradict the value of the MSE.
This is particularly evident for FC topologies with SL coupling in the transient regime, which display a lower averaged $\chi$ and at the same time a lower MSE.
This apparent paradox can be traced back to this data reporting only the \textit{averaged} Holevo $\chi$. 
To gain further insight into this phenomenon we report in~\cref{fig:holevo single nodes} the time-dependence of $\chi$ corresponding to each node, averaging only over the three input ensembles discussed in~\cref{sec:holevo}.
This data shows that in the transient regime information can be asymmetrically distributed across the reservoir nodes.
In particular, $\chi$ tends to be much higher for reservoir nodes close to the input.
Having few such nodes with a high $\chi$ and many other nodes with vanishingly small $\chi$ results in a relatively large average $\chi$ even in situations where the number of reservoir qubits correlated to the input is not sufficient for state reconstruction.
On the other hand, for ML-coupled systems information spreads much more symmetrically, and thus also the average $\chi$ is inversely correlated with the MSE, as expected.
Beyond the scrambling time every single $\chi$ saturates to the same value of $\chi \simeq (2.5\pm1.2)\cross10^{-3}$, as expected from~\cref{fig:General Results}.

Overall, all the reported data shows a robust convergence to a regime where state estimation is possible with the same performances granted by random unitary dynamics, without the need to fine tune the interaction times.
{The initial differences between interaction topologies --- attributable to different rates of information spreading --- gradually converge to a uniform behaviour over longer timescales, once information has spread evenly across the network.}
Furthermore, if such fine-tuning is feasible, even higher performances are possible for some types of interactions in the transient regime before the scrambling time.

\section{Discussion}
\label{sec:conclusions}

We provided strong evidence that QELM-based accurate reconstruction~\cite{innocenti_potential_2023,suprano_experimental_2023} with local measurements is possible well beyond the scrambling time for several types of dynamics.
In fact, we showed that for such evolution times, the reconstruction performance is identical to the one obtained with Haar-random local unitaries, which result in the maximum amount of distributed correlation across a system.
These findings are interesting for both their experimental implications, and for the insights they provide into the relations between QELMs and QIS.

From an experimental perspective, our findings mean that for many types of dynamics there is no need to fine-tune the evolution time for the purpose of reconstructing properties of input states via QELMs.
As long as the system is left to evolve long enough for the information to spread uniformly, accurate reconstruction is always possible. The evolution time must still remain fixed across different measurements, but its precise value does not need to be known to the experimenter.

At a more fundamental level, our findings offer a novel perspective into the nature of QIS and its relations to QELMs and state reconstruction tasks.
Even though a scrambling system is considered to be one where information is hidden in the correlation and is locally irretrievable, our results highlight that --- at least for relatively small systems --- the opposite is true from a state estimation perspective: when information is left to spread uniformly throughout the reservoir qubits, the residual local information is still sufficient to retrieve arbitrary properties of input states.

In summary, our results pave the way for robust experimental state reconstruction schemes that do not rely on accurate knowledge of the underlying dynamic or fine-tuning of the experimental apparatus, and furthermore suggest that the way information spreads locally for scrambling system might be a useful resource for quantum state estimation purposes.

\section{Methodology}
\label{sec:task}

In this section, we will give the technical details on how we generated the tested dynamics and computed reconstruction MSE and QIS.

As reservoirs, we consider $(N+1)$-qubit systems undergoing a time-independent unitary evolution, with the first $N$ qubits making up the reservoir, and the ``\textrm{in}'' qubit accommodating the input state.
The overall evolution is generated by the Hamiltonian $H=H_{\rm res}+H_{\rm inj}$, with $H_{\rm res}$ the interaction between the reservoir qubits, and $H_{\rm inj}$ the interaction between input and reservoir.
The interaction between reservoir qubits is modelled by a general Hamiltonian of the form
\begin{equation}
        H_{\rm res} = \sum^{3}_{\alpha,\beta=1}\sum^N_{k<j}J^{\alpha, \beta}_{k,j}\sigma^\alpha_k\otimes \sigma^\beta_j + \sum^{3}_{\alpha=1}\sum^{N}_{k=1}\Delta^\alpha_k\sigma^\alpha_k,
        \label{eq:hamiltonian_res}
\end{equation}
where the indices $k,j=1,...,N$ run over the reservoir qubits, and $\{\sigma^\alpha_j\}_{\alpha=1}^3$ are the Pauli matrices acting on the $j$-th-qubit.
To study how different topologies affect estimation accuracy, we consider systems where the reservoir qubits are arranged in a chain (C), a ring (R), or are fully connected (FC).
The interaction between input and reservoir has the form
\begin{equation}
    H_{\rm inj} = \sum^{3}_{\alpha=1}\sum^N_{k=1}J^{\alpha, \beta}_{k,\rm in}\sigma^\alpha_k\otimes \sigma^\beta_{\rm in}.
    \label{eq:hamiltonian_injection}
\end{equation}
We consider two types of interaction between input and reservoir:
a ``single link'' (SL) coupling scheme ($J^{\alpha,\beta}_{k,\rm in} = J^{\alpha,\beta}_{k,\rm in}\delta_{1,k}$ , where the input qubit interacts with a single reservoir qubit, and a ``multi link'' (ML) coupling scheme where the input directly interacts with all reservoir qubits.
These different interaction topologies are summarized in~\cref{fig:configurazioni}.
In each case, the interaction parameters $J_{ij}^{\alpha,\beta}$ and $\Delta_i^\alpha$ are sampled uniformly at random in the range $[-1,1]$ and $[-0.1,0.1]$, respectively.
For each topology, we tested $500$ random Hamiltonians. Unless differently specified, we considered a reservoir comprised of $N=7$ qubits.
The sampling statistics is fixed to $10^6$ in each case.
As the MSE is known to scale as $1/N$ with $N$ the sampling statistics, there is no loss of generality in fixing this value, since we are only interested in the topology effects on the performances.

For QELMs we focus on the on a single qubit state tomography, that is, we fix $\Ytrain_{ij}=\Tr[\sigma_i \rho_j^{\rm train}]$, from the estimated expectation values of $\sigma_z$ on the reservoir qubit after the evolution.

To relate this with QIS, we then compute the OTOC using~\cref{eq: C otoc} with $\mathbf O_A=\sigma^z_i$, $i=1,...,N$, and $\mathbf{O}_B=\sigma^\alpha_{\rm in}$, $\alpha=1,2,3$ as input node operators:
\begin{equation}
    C^i_\alpha(t) = 1 -\frac{1}{2^N}\Tr{\sigma^z_i(t)\sigma^\alpha_{\rm in}\sigma_i^z(t)\sigma^\alpha_{\rm in}}
    \label{eq:cialfa case study}
\end{equation}
We will then consider the average of this quantity over $i$ and $\alpha$~\cite{landsman_verified_2019}, to quantify the average correlation between the output local measurements and target observables we mean to reconstruct:
\begin{equation}
    C(t) = 1 - \frac{1}{3N}\frac{1}{2^N}\sum^{3}_{\alpha=1}\sum^N_{i=1}C^i_\alpha
    \label{eq:commutator case study}
\end{equation}

Finally, following the notation in~\cref{sec:holevo}, we compute the average Holevo information fixing as input states the eigenstates of the three Pauli matrices.
More explicitly, we compute the Holevo information with respect to input ensembles $\{(p_1,\rho^1),(p_2,\rho^2)\}$ with $p_i=1/2$ and $\rho^1, \rho^2$ the eigenstates of one of $\sigma_x$, $\sigma_y$, and $\sigma_z$.
As dynamical maps we use $\Phi_t^{(k)}(\rho)=\Tr_{\bar k}[e^{-iHt} \rho e^{iHt}]$ with $\Tr_{\bar k}$ denoting the trace over all but the $k$-th qubit.
These choices ensure we obtain a quantity which directly relates to the amount of information about the expectation values of $\sigma_x,\sigma_y,\sigma_z$ retrievable from local measurements on the reservoir.
We finally compute the average for each $k$, obtaining a local Holevo information corresponding to each reservoir qubit.
The resulting quantifier thus provides an easily computable upper bound for the correlations between the input state at $t=0$ and the $j$-th output qubit at time $t$.

\section*{Data Availability}

All relevant data and figures supporting the main conclusions of the document are available on request. Please refer to Marco Vetrano at marco.vetrano02@unipa.it.

\section*{Code Availability}

All relevant code supporting the document is available upon request. Please refer to Marco Vetrano at marco.vetrano02@unipa.it.

\begin{acknowledgments}
MV and GMP acknowledge funding under project PNRR - Research infrastructures: Strengthening the Italian leadership in ELT and SKA (STILES). 
GLM and GMP acknowledge funding from the European Union - NextGenerationEU through the Italian Ministry of University and Research under PNRR - M4C2-I1.3 Project PE-00000019 "HEAL ITALIA" (CUP B73C22001250006 ). 
SL and GMP  acknowledge support by MUR under PRIN Project No. 2022FEXLYB.
Quantum Reservoir Computing (QuReCo) and by the “National Centre for HPC, Big Data and Quantum Computing (HPC)” Project CN00000013 HyQELM – SPOKE 10.
LI acknowledges support from MUR and AWS under project PON Ricerca e Innovazione 2014-2020, ``Calcolo quantistico in dispositivi quantistici rumorosi nel regime di scala intermedia" (NISQ - Noisy, Intermediate-Scale Quantum). 

\end{acknowledgments}

\section*{Contributions}

MV developed the code with the contribution of SL. SL and GMP developed the initial idea. All  authors contributed to write the manuscript. All authors contributed to the discussions and interpretations of the results.

\section*{Competing Interests}

All authors declare no financial or non-financial competing interests.

\bibliography{bibliography}

\end{document}